\documentclass[11pt,a4paper]{article} 
\pdfoutput=1
\usepackage{adjustbox}
\usepackage{jheppub}
\usepackage{enumerate}
\usepackage{epsfig} 
\usepackage{float} 
\usepackage[caption = false]{subfig}
\usepackage{wasysym} 
\usepackage{mathrsfs} 
\usepackage{amsfonts} 
\usepackage{amsbsy} 
\usepackage{amscd} 
\usepackage{pstricks} 
\usepackage{multirow} 
\usepackage{tikz}
\usepackage{color}
\usepackage{slashed}
\usepackage{array}
\usepackage{tablefootnote}
\usepackage{amsmath}
\usepackage{orcidlink}

\usetikzlibrary{arrows,positioning,shapes.geometric} 
\usepackage[compat=1.1.0]{tikz-feynman}          % package allowing Feynman diagrams
\usepackage[font=small,labelfont=bf]{caption}
\usepackage{slashed}
\usepackage{multirow}
\usepackage{tabularx}
\usepackage{soul}  % Strikethrough text with \st{Hellow world}
\usepackage{listings} 
\usepackage{color}
\usepackage{amsthm}

\newcommand{\et}{$\mathbf{E(2)}$}
\newcommand{\ot}{$\mathbf{O(2)}$}
\title{
Equivariant, Safe and Sensitive \\\hfill --- Graph Networks for New Physics}
\abstract{This study introduces a novel Graph Neural Network (GNN) architecture that leverages infrared and collinear (IRC) safety and equivariance to enhance the analysis of collider data for Beyond the Standard Model (BSM) discoveries. By integrating equivariance in the rapidity-azimuth plane with IRC-safe principles, our model significantly reduces computational overhead while ensuring theoretical consistency in identifying BSM scenarios amidst Quantum Chromodynamics backgrounds. The proposed GNN architecture demonstrates superior performance in tagging semi-visible jets, highlighting its potential as a robust tool for advancing BSM search strategies at high-energy colliders.}
\author[a]{Akanksha Bhardwaj\orcidlink{0000-0001-8695-972X},}
\author[b]{Christoph Englert\orcidlink{0000-0003-2201-0667},}
\author[b]{Wrishik Naskar\orcidlink{0000-0002-4357-899},} 
\author[c]{Vishal S. Ngairangbam\orcidlink{0000-0002-7143-715X},}
\author[c]{and Michael Spannowsky\orcidlink{0000-0002-8362-0576}}
\affiliation[a]{Department of Physics, Oklahoma State University, Stillwater, OK, 74078, USA}
\affiliation[b]{School of Physics and Astronomy, University of Glasgow, Glasgow G12 8QQ, United Kingdom}
\affiliation[c]{Institute for Particle Physics Phenomenology, Department of Physics,\\ Durham University, Durham DH1 3LE, United Kingdom}
\emailAdd{akanksha.bhardwaj@okstate.edu}
\emailAdd{christoph.englert@glasgow.ac.uk}
\emailAdd{w.naskar.1@research.gla.ac.uk}
\emailAdd{vishal.s.ngairangbam@durham.ac.uk}
\emailAdd{michael.spannowsky@durham.ac.uk}

\preprint{ IPPP/24/07}

\allowdisplaybreaks

\begin{document}
\maketitle
\flushbottom
%%%%%%%%%%%%%%%%%%%%%%%%%%%
\section{Introduction}
%%%%%%%%%%%%%%%%%%%%%%%%%%%
The application of machine learning algorithms to jet classification~\cite{Cogan:2014oua,deOliveira:2015xxd,Kasieczka:2017nvn,Louppe:2017ipp,Dreyer:2018nbf,Heimel:2018mkt,Farina:2018fyg,Qu:2019gqs,Kasieczka:2019dbj,Bernreuther:2020vhm,pmlr-v162-qu22b,Kasieczka:2020nyd,Dreyer:2020brq,Khosa:2021cyk,Canelli:2021aps,Dreyer:2021hhr,Cavallini:2021vot,Butter:2022xyj,Dillon:2022mkq,Faucett:2022zie,Keicher:2023mer,Finke:2023veq,Lu:2023gjk,Bardhan:2023mia,Metodiev:2023izu,Favaro:2023xdl,Romero:2023hrk,Furuichi:2023vdx,Gaertner:2023ycs} provides an ideal environment to gauge the interplay between performance and interpretability. On the one hand, we need highly performant algorithms utilising the wealth of experimental data recorded at the Large Hadron Collider (LHC), while on the other, we want to ascertain the reason behind the algorithm's efficiency. Architectures which are Lorentz equivariant~\cite{Bogatskiy:2020tje,Gong:2022lye,Li:2022xfc,Hao:2022zns,Bogatskiy:2022czk}, or are infra-red and collinear (IRC) safe~\cite{Choi:2018dag,Komiske:2018cqr,Konar:2021zdg,Atkinson:2022uzb,Athanasakos:2023fhq,Shen:2023ofd,Konar:2023ptv,Bright-Thonney:2023gdl} have been shown to enhance the physical biases of the algorithms with generally minimal loss, or at times, comparatively increased performance~\cite{Gong:2022lye,Bogatskiy:2022czk} with respect to physics-opaque constructions. In this work, we combine IRC safety with Euclidean equivariance in the rapidity-azimuth plane. With well-known examples of jet-shape~\cite{Almeida:2008yp,Thaler:2010tr,Thaler:2011gf,Larkoski:2013eya,Larkoski:2014pca,Moult:2016cvt,Dasgupta:2015lxh,Caletti:2021oor} variables, constructed as functions of distance in the rapidity-azimuth plane, this pushes the IRC-safe features closer to human-engineered QCD features without losing out on performance, even with two orders of magnitude lower number of model parameters.   

Symmetry in a different sense, namely the degeneracy over nearly degenerate final states, is fundamental to the theoretical robustness of information that can be gained from phenomenological analyses. The Kinoshita-Lee-Nauenberg theorem~\cite{Kinoshita:1962ur,Lee:1964is} guarantees cancellations of soft and (final-state) collinear divergences when algorithms are infrared (IR) and collinear safe; any departure from IRC-safe methods ultimately implies uncorrectable interpretation shortfalls. This technically elevates IRC safety to a comparable level as the space-time symmetries to perturbatively model and interpret scattering processes. 

The recent surge in the application of machine learning for enhancing the sensitivity to new physics in available and projected particle physics data sets has led to a departure from observables that transparently reflect symmetry properties and IRC safety. Nonetheless, any such serious attempt should critically feature IRC safety whilst exploiting symmetries of the data sets in the corresponding architecture. This work aims to demonstrate that this is not only feasible but directly applicable to the determination of BSM parameter regions in case a discovery can be claimed via these techniques. We construct an IRC-safe \et-equivariant network and demonstrate the enhanced optimisation capability of such an algorithm compared to approaches that do not exploit symmetry.   As most popular substructure variables are generally built out of $\Delta R=\sqrt{(\Delta y)^2+(\Delta\phi)^2 }$, our choice of the \et-group preserves such a structure in the feature extraction process, thereby closely matching the QCD-intuitive picture of high-level variables. We then apply this approach to a classification of jet substructure analyses (we consider Hidden Valley Models~\cite{Strassler:2006im,Carloni:2010tw,Carloni:2011kk,Cohen:2015toa,Kar:2020bws,Bernreuther:2020xus,Barron:2021btf,Albouy:2022cin} as a representative example), where IRC safety plays a critical role in obtaining a theoretically robust output score when the substructure becomes sparse or~soft.

This work is organised as follows: In Sec.~\ref{sec:O2equivariant}, we discuss the inductive biases relevant to phenomenology at the LHC, namely IRC safety and equivariance in the rapidity-azimuthal plane.  Then, in Sec.~\ref{sec:application}, we apply this network to a relevant physics case where it discriminates between QCD jets and dark shower jets (semi-visible jets). We discuss the network architecture and its performance. Finally, we provide a summary and conclusion in Sec.~\ref{sec:Summary}.

%%%%%%%%%%%%%%%%%%%%%%%%%%%
\section{Inductive biases and QCD}
\label{sec:O2equivariant}
%%%%%%%%%%%%%%%%%%%%%%%%%%%
The success of deep learning algorithms relies on reducing the search space of possible functions to a predetermined subset by imposing additional structures on the architecture. These inductive biases favour the extraction of relevant features which are not too specific (like high-level variables) but follow the intuition of the application domain. Here, we discuss the two inductive biases we impose on the Message Passing Neural Network utilised in our study: IRC safety and equivariance in the rapidity-azimuth plane.  

%%%%%%%%%%%%%%%%%%%%%%%%%%%
\subsection{Infrared and Collinear Safety} 
%%%%%%%%%%%%%%%%%%%%%%%%%%%
From a QCD perspective, infra-red and collinear safe feature extraction promises to provide better interpretability while not losing out on the high classification capabilities.  Any message passing operation involves taking in the node features $\mathbf{H}^{(l)}_i$ of each node $i$ as the input  and updating it to $\mathbf{H}^{(l+1)}_i$.  As IRC safety is defined at the level of the whole jet and not for each constituents,  these node features inevitably undergo a graph-readout operation for jet-level classification. A way to define an IRC-safe graph-representation is the ``\emph{conservation of collinearity}" of the node features  in each representation indexed by $l$ of the message passing operation which preserves the structure of the angular components  in the limit of collinear emissions.  In the following, each particle $i$ has four vectors $p^\mu_i=(z_i,\hat{\mathbf{p}}_i)$, with  $z_i=p^i_T/\sum_k p^k_T$ and $\hat{\mathbf{p}}_i$ denoting the angular vector. Mathematically, a node representation $\mathbf{H}^{(l)}_i$ conserves collinearity, if for any particle $q$, $r$, and $s$, we have 
	\begin{equation} 
		\label{eq:cons_col} 
		z_q=z_r+z_s \land \hat{\mathbf{p}}_q=\hat{\mathbf{p}}_r=\hat{\mathbf{p}}_s \implies \mathbf{H}^{(l)}_q=\mathbf{H}^{(l)}_r=\mathbf{H}^{(l)}_s\quad. 
	\end{equation}  
Once this condition is satisfied, it is straightforward to define an IRC safe graph representation as 
\begin{equation}
	\label{eq:irc_graph_read} 
	\mathbf{G}_l=\sum_i z_i\;\mathbf{H}_i^{(l)}\quad. 
	\end{equation}
	
For implementing IRC safety in Energy-weighted Message Passing Network (EMPN)~\cite{Konar:2021zdg}, one defines an IRC-safe prescription for defining the graph structure. Once this is satisfied for any node $i$ in the graph,
 the message-passing operation
\begin{equation}
	\label{eq:ewmp}
	\mathbf{H}^{(l+1)}_i=\sum_{j\in\mathcal{N}[i]}\;\omega^{(\mathcal{N}[i])}_{j}\;\;\hat{\Phi}^{(l+1)}(\mathbf{H}^{(l)}_i,\mathbf{H}^{(l)}_j)\quad
\end{equation} 
where the scope-dependent factors are an analogue of $z_i$, defined as 
\begin{equation*}
	\omega_j^{(\mathcal{N}[i])}=\frac{p_T^j}{\sum_{k\in\mathcal{N}[i]}\;p_T^k} \quad,
\end{equation*}
recursively conserves collinearity of the node features if the initial input $\mathbf{H}^{(0)}_i$ follows the same. For the inpuit, one can use the relative coordinates in the rapidity-azimuth plane from the jet axis, $\mathbf{H}^{(0)}_i=(\Delta y_{iJ},\Delta\phi_{iJ})$ satisfies the condition trivially.   
The function $\hat{\Phi}^{(l)}$, in general, can be any well-behaved neural network which we take to be an edge convolution network~\cite{DBLP:journals/corr/abs-1801-07829}. The neighbourhood $\mathcal{N}[i]$ is constructed by imposing the step function $\Theta(\Delta R_{ij}<R_0)$ for each particle $i$, $R_0$ being a parameter that determines the density of the constructed graph. 
The graph features (by taking a graph readout as defined in Eq.~\ref{eq:irc_graph_read}) for $l>0$ can be concatenated up to the total number of such operations $L$, which is a hyperparameter. For classification, a downstream network takes the graph representation as an input and gives a classification score. 

%%%%%%%%%%%%%%%%%%%%%%%%%%%
\subsection{Equivariance in rapidity-azimuth plane}  
%%%%%%%%%%%%%%%%%%%%%%%%%%%
Equivariance is at the core of the powerful learning ability shown by modern deep-learning algorithms such as Convolutional Neural Networks (CNNs) and Graph Neural Networks (GNNs). They reduce the possible set of functions that a neural network can approximate by assuming symmetries of the input data. For instance, CNNs assume translational symmetry in $\mathbb{R}^2$ while GNNs assume permutation symmetry of the nodes. Equivariance enforces group algebraic structures in the hidden representations in a neural network. This reduces the set of possible functions that a neural network can approximate to those which follow the group equivariant property.  

Mathematically, for a group $\mathcal{G}$ with an element denoted as $g$, let $\mathbf{X}$ and $\mathbf{Y}$ be two sets with $\mathbf{x}\in\mathbf{X}$ and $\mathbf{y}\in\mathbf{Y}$ that admit group actions\footnote {An action of a group $\mathcal{G}$ is defined as $\mathbf{T}_\mathbf{X}:\mathcal{G}\times\mathbf{X}\to\mathbf{X}$ such that for any $\mathbf{x}\in\mathbf{X}$, we have $\mathbf{T}_\mathbf{X}(e,\mathbf{x})=\mathbf{x}$ for the neutral element $e\in \mathcal{G}$ and $\mathbf{T}_\mathbf{X}(g_1,\mathbf{T}_\mathbf{X}(g_2,\mathbf{x}))=\mathbf{T}_\mathbf{X}(g_1g_2,\mathbf{x})$, for all $g_1,g_2\in\mathcal{G}$.} $\mathbf{T}_\mathbf{X}(g,\mathbf{x})$ and $\mathbf{T}_\mathbf{Y}(g,\mathbf{y})$, respectively. A function $f:\mathbf{X}\to\mathbf{Y}$ is $\mathcal{G}$-equivariant if
\begin{equation}
	\label{eq:g_equiv} 
	f(\mathbf{T}_\mathbf{X}(g,\mathbf{x}))=\mathbf{T}_\mathbf{Y}(g,f(\mathbf{x}))\quad. 
\end{equation}  
An equivariant neural network~\cite{Lim2023} approximates such a function $f$ for a given group, and $f$ is generally non-linear. From an approximation perspective, the neural network being able to learn only those functions that follow Eq.~\eqref{eq:g_equiv} rather than any general function necessarily reduces the expressive power. However, the reduction generally results in higher efficiency of the optimisation process in terms of network complexity, on top of increasing the interpretability as the particular group $\mathcal{G}$ is determined from domain knowledge.   
If the action $\mathbf{T}_\mathbf{Y}$ is trivial, i.e., $\mathbf{T}_\mathbf{Y}(g,\mathbf{y})=\mathbf{y}\;\forall g\in\mathcal{G}$, then the function $f$ is said to be $\mathcal{G}$-invariant.  GNNs are permutation equivariant in the node update stage, while for graph-level purposes, the global readout renders the graph representation permutation invariant.  

For our purpose, the group which preserves the separation $\Delta R$ in the $(y,\phi)$ plane is the Euclidean group $\mathbf{E}(2)$. To incorporate $\mathbf{E}(2)$-equivariance in an EMPN, we modify the general $\mathbf{E}(n)$-equivariant message-passing operation of reference~\cite{pmlr-v139-satorras21a}.  To define   $\mathbf{E}(n)$ equivariance in this form, one segregates the node features $\mathbf{H}^{(l)}_i$, into a scalar and a vector representation. Denoting the former as $\mathbf{h}^{(l)}_i$ and the latter as $\mathbf{x}^{(l)}_i$, the $\mathbf{E}(n)$ equivariant operation is given as 
\begin{subequations}
	\label{eq:en_equi} 
		\begin{align}  
			\label{eq:en_equi_mij}
		\mathbf{m}^{(l+1)}_{ij}&=\Phi^{(l+1)}_e(\mathbf{h}^{(l)}_i,\mathbf{h}^{(l)}_j,|\mathbf{x}^{(l)}_i-\mathbf{x}^{(l)}_j|)\quad,\\
		\label{eq:en_equi_xi}
		\mathbf{x}_i^{(l+1)}&=\mathbf{x}^{(l)}_i+\sum_{j\neq i} \,\; (\mathbf{x}^{(l)}_i-\mathbf{x}^{(l)}_j) \; \Phi^{(l+1)}_x(\mathbf{m}^{(l+1)}_{ij})\quad,\\
		\label{eq:en_equi_mi}
		\mathbf{m}^{(l+1)}_i&=\sum_{j\neq i}\,\;\mathbf{m}^{(l+1)}_{ij}\quad,\\
		\label{eq:en_equi_hi}
		\mathbf{h}^{(l+1)}_i&=\Phi_h(\mathbf{h}^{(l)}_i,\mathbf{m}^{(l+1)}_i) \quad,
	\end{align} 
\end{subequations} 
where ${\Phi}^{(l+1)}_e$, ${\Phi}^{(l+1)}_x$, and ${\Phi}^{(l+1)}_h$  are multilayer perceptrons (MLPs), with $\Phi^{(l+1)}_x$ giving a one-dimensional output with a sigmoid activation, thereby being interpreted as a scalar weight in the aggregation.
 The restriction $j\neq i$ in the aggregation steps can be relaxed as the equivariance is satisfied in its absence. Moreover, as the neighbourhood construction modifies the message functions (${\Phi}^{(l+1)}_e$ and ${\Phi}^{(l+1)}_x$) by a product with a step function $\Theta(\Delta R_{ij}<R_0)$ which is itself $\mathbf{E}(2)$ invariant, we can incorporate a local aggregation scheme without impeding global equivariance. We note that modifying Eq.~\eqref{eq:ewmp} to 
\begin{equation}
	\label{eq:ang_func_irc}
     		\mathbf{H}^{(l+1)}_i=\mathbf{F}^{(l+1)}\left(\mathbf{H}_i^{(l)},\sum_{i\in\mathcal{N}[i]}\;\omega^{(\mathcal{N}[i])}_{j}\;\;\hat{\Phi}^{(l+1)}(\mathbf{H}^{(l)}_i,\mathbf{H}^{(l)}_j)\right)\,,
\end{equation} 
with a well-behaved (to be learned) function $\mathbf{F}^{(l+1)}$ still conserves collinearity of the updated node features. This is because the conservation of collinearity in the second argument already requires its conservation in the first argument $\mathbf{H}^{(l)}_q=\mathbf{H}^{(l)}_r=\mathbf{H}^{(l)}_s$. In general, by the definition of a function not being one-to-many, any function $f(\mathbf{H}^{(1)}_i,\mathbf{H}^{(2)}_i,...,\mathbf{H}^{(n)}_i)$ of $n$ node representations $\mathbf{H}^{(l)}_i$ of a single node feature $i$ conserves collinearity, if each $\mathbf{H}^{(l)}_i$ conserves collinearity. Equipped with this observation, it is straightforward to modify  Eq.~\eqref{eq:en_equi} to satisfy IRC safety by incorporating the energy-weighted sum structure in the aggregation as
\begin{subequations} 
\label{eq:irc_en_equiv}
\begin{align}
		\label{eq:irc_en_equiv_mij}
		\mathbf{m}^{(l+1)}_{ij}&=\Phi^{(l+1)}_e(\mathbf{h}^{(l)}_i,\mathbf{h}^{(l)}_j,|\mathbf{x}^{(l)}_i-\mathbf{x}^{(l)}_j|^2)\quad,\\
		\label{eq:irc_en_equiv_xi}
		\mathbf{x}_i^{(l+1)}&=\mathbf{x}^{(l)}_i+\sum_{j\in\mathcal{N}[i]} \omega^{(\mathcal{N}[i])}_j\; (\mathbf{x}^{(l)}_i-\mathbf{x}^{(l)}_j) \; \Phi^{(l+1)}_x(\mathbf{m}^{(l+1)}_{ij})\quad,\\
		\label{eq:irc_en_equiv_mi}
		\mathbf{m}^{(l+1)}_i&=\sum_{j\in\mathcal{N}[i]} \omega^{(\mathcal{N}[i])}_j\;\mathbf{m}^{(l+1)}_{ij}\quad,\\
		\label{eq:irc_en_equiv_hi}
		\mathbf{h}^{(l+1)}_i&=\Phi^{(l+1)}_h(\mathbf{h}^{(l)}_i,\mathbf{m}^{(l+1)}_i)  \quad. 
\end{align}
\end{subequations}  
Conservation of collinearity of the updated scalar features $\mathbf{h}^{(l+1)}_i$, and vector features $\mathbf{x}^{(l+1)}_i$ follows from comparing their corresponding aggregation equations to Eq.~\eqref{eq:ang_func_irc}. We have modified the input Euclidean norm to the squared value as we have self-loops, and the square root is not differentiable at zero. 

The general $\mathbf{E}(n)$ group consists of translations and rotations. While there are some instances where the translational symmetry is utilised by defining vector observables like jet pull~\cite{Gallicchio:2010sw,ATLAS:2018olo,Larkoski:2019fsm},  most substructure observables are confined to a rotationally symmetric definition by using only the scalar separation $\Delta R$.  Therefore, we also study an $\mathbf{O}(2)$-equivariant architecture by forgoing translational equivariance. This can be done by modifying Eqs.~\eqref{eq:irc_en_equiv_mij} and \eqref{eq:irc_en_equiv_xi} to
\begin{subequations}
	\label{eq:irc_on_equiv}
	\begin{align}
		\label{eq:irc_on_equiv_mij}  
	\mathbf{m}^{(l+1)}_{ij}&=\Phi^{(l+1)}_e(\mathbf{h}^{(l)}_i,\mathbf{h}^{(l)}_j,|\mathbf{x}^{(l)}_i-\mathbf{x}^{(l)}_j|^2,\mathbf{x}^{(l)}_i.\,\mathbf{x}^{(l)}_j)\quad,\\
	\label{eq:irc_on_equiv_xi}
	\mathbf{x}_i^{(l+1)}&=\sum_{j\in\mathcal{N}[i]} \omega^{(\mathcal{N}[i])}_j\;\mathbf{x}^{(l)}_j \; \Phi^{(l+1)}_x(\mathbf{m}^{(l+1)}_{ij})\quad,
	\end{align} 
\end{subequations} 
and keeping Eqs.~\eqref{eq:irc_en_equiv_mi} and \eqref{eq:irc_en_equiv_hi} unchanged.  We have added the inner product $\mathbf{x}^{(l)}_i.\,\mathbf{x}^{(l)}_j$ as it is invariant to rotations, which is not the case for translations.    

\begin{figure}[!t]
	\centering
	\includegraphics[width=1\linewidth]{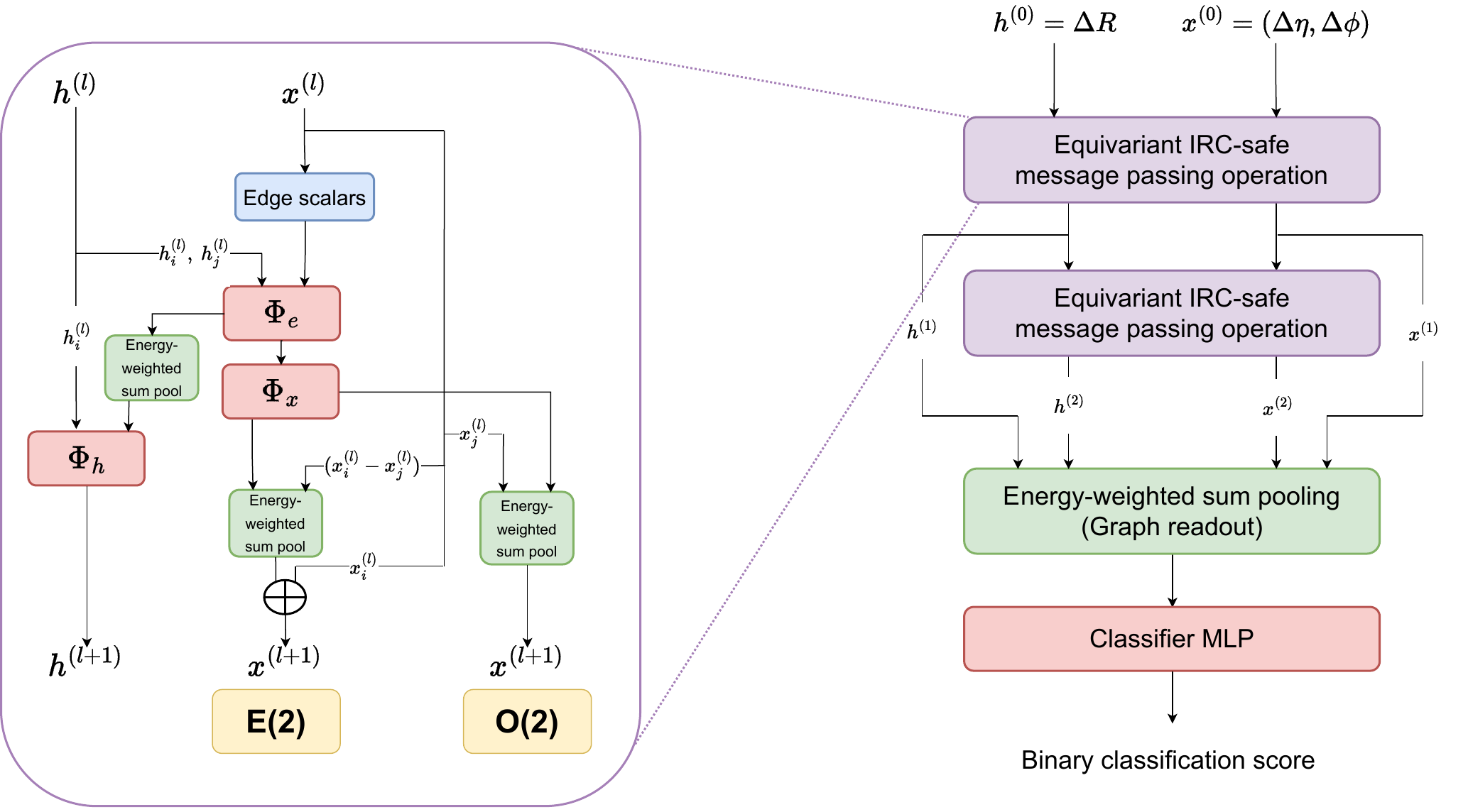}
	\caption{A schematic representation of $\mathbf{E}(2)$-EMPN and $\mathbf{O}(2)$-EMPN network architecture. \label{fig:o2_architecture}}
\end{figure}

Note that for the final IRC safe feature extraction, we need to perform an energy-weighted summed global readout as given in Eq.~\eqref{eq:irc_graph_read} for both $\mathbf{x}^{(l)}_i$ and $\mathbf{h}^{(l)}_i$. Moreover, the input scalars $\mathbf{h}^{(0)}_i$ cannot contain $z_i$ from the requirement of IRC safety, even though it is invariant under $\mathbf{E}(2)$.  A schematic representation of the equivariant message passing is shown on the left  of figure~\ref{fig:o2_architecture}. 

%---------------------
%---------------------

%%%%%%%%%%%%%%%%%%%%%%%%%%%
\section{A Physics application: Tagging semi-visible jets}
%%%%%%%%%%%%%%%%%%%%%%%%%%%
\label{sec:application}
The radiation pattern arising from perturbative QCD emissions is known to be approximately uniform in the rapidity-azimuth plane. This guides the design of most jet-shape observables~\cite{Almeida:2008yp} to use the relative distance between $\Delta R$, between pairs of particles~\cite{Larkoski:2013eya,Moult:2016cvt}, or from well-defined axes~\cite{Thaler:2010tr,Thaler:2011gf}. The implicit assumption underlying such definitions is the invariance of the observables under the two-dimensional Euclidean group $\mathbf{E}(2)$ in the rapidity-azimuth plane. It is, therefore, a natural extension of IRC-safe feature extraction to incorporate $\mathbf{E}(2)$-equivariance to better connect to known jet substructure observables in literature.

Although equivariance, in principle, would help segregate QCD radiation patterns with different $n$-prong structures, a particularly challenging case would be when the parton shower itself is altered with non-trivial physics. For example, a QCD-like dark shower in Hidden Valley models can produce a mixture of BSM and normal SM particles, generally modelled via interleaved parton showers. In such a case, the differences that arise from the QCD radiation are not in the presence of the hard prongs but rather at an intricate level in the relative distribution of the particle's position in the $(y,\phi)$ plane, with possibly some missing patches left by undetectable stable or quasi-stable particles within the jet. Generally, the energy values would be similar to a one-prong QCD jet. At the same time, the relative position of the particles in the $(y,\phi)$ plane would be the distinguishing feature in such semi-visible jets. Therefore, $\mathbf{E}(2)$ equivariance would be more prominent in identifying such semi-visible jets. 

Hidden Valley Models offer an intriguing perspective in particle physics, suggesting the existence of secluded, almost decoupled sectors that interact weakly with the SM. Such a scenario can be brought about by extending the SM gauge group with a new group $G_v$. The choices of the gauge group and new fermions and their relation to the SM particles offer a rich phenomenology with possibly distinct signatures like emerging jets or semi-visible jets not covered by traditional searches at the LHC, like monojet searches~\cite{CMS:2021far,ATLAS:2022ckd}.  In general, the  SM particles remain neutral under the new group $G_v$, while there can be two classes of new particles: those charged under the SM gauge group and the $G_v$ and those which solely have a $G_v$ charge.  Interactions between the new sector and SM particles can occur through intermediary mechanisms, such as a TeV scale $Z'$ or through higher-dimensional operators and loops involving heavy particles possessing non-trivial $G_{\text{SM}}$ and $G_v$ charges. The case of a confined dark sector decoupled from SM in the lower energies is of particular interest.

%%%%%%%%%%%%%%%%%%%%%%%%%%%
\subsection{Simulation details}
%%%%%%%%%%%%%%%%%%%%%%%%%%%
Strongly coupled hidden-sector particles produced at the LHC could produce invisible dark mesons along with stable mesons. In such a case, novel LHC signatures such as semi-visible jets could be produced, almost indistinguishable from QCD backgrounds, therefore posing an exciting challenge for jet classification. This study considers  resonant and non-resonant production of semi-visible jets. For the latter, we consider both Electroweak (EW) and QCD productions separately.  We simulate these dark shower jets events at 14 TeV centre-of-mass energy proton-proton collisions using the Hidden Valley (HV) module of \textsc{Pythia8.310}~\cite{Bierlich:2022pfr}.  They are generated as: 
\begin{itemize}
\item \textbf{Resonant production:} A new leptophobic vector boson $Z'$ is produced as a BSM mediator, decaying to a pair of dark quarks. Events are generated setting with the process \texttt{HiddenValley:ffbar2Zv = on}. We set the mass of $Z'$ = 1 TeV with a width of 10 MeV. The $Z'$ decays to a pair of dark quarks ($q_d\,\bar{q}_d)$, which has no SM charge but can decay to another dark sector particle and an SM quark. The subsequent showering and hadronisation produce dark mesons and SM particles.
\item \textbf{Non-resonant EW production:} We produce a pair of $U_v$ quarks using the process \texttt{HiddenValley:ffbar2UvUvbar = on}. These $U_v$'s are charged under the fundamental representation of the dark SU(N) and mirror the SM charges of the $u$-quark. It, therefore, radiates into both SM and dark sector particles.
\item \textbf{Non-resonant QCD production:} The same $U_v$ pair of quarks are produced using the process  \texttt{HiddenValley:gg2UvUvbar = on}. Unlike the previous channel, which occurs via an s-channel colour singlet propagator, this channel has an SM colour connection between the initial and the final state lines, resembling QCD dijet productions more closely.  
\end{itemize}
The fraction of stable to total dark mesons produced in the process is controlled by the \texttt{probVector} parameter in the HV Module. We have chosen \texttt{probVector}=0.5 in our analysis. This choice is motivated by higher values producing events with low missing transverse energy, which is covered by dijet bump hunts. In comparison, lower values correspond to signatures where the missing transverse energy's $\phi$ direction is away from the jet axes and, hence, are covered by more traditional dark matter searches. 

All final state particles detectable at the LHC are clustered into microjets of radius $R=0.1$ using the anti-$k_t$ algorithm  \cite{Cacciari:2008gp}, mimicking the calorimeter resolution. Inclusive microjets with transverse momentum $p_T>1$ GeV are then clustered into jets of radius $R=0.8$ with the anti-$k_t$ algorithm. The jet clusterings are performed using the \textsc{FastJet (v3.4.1)}~\cite{Cacciari:2011ma} package.  Events are required to have at least two jets within $|\eta| < 3$, the leading jet is required to have $p_T >  150~\text{GeV}$, and other sub-leading jets are required to have $p_T >  120~\text{GeV}$. 

 For the background, QCD dijet events are generated by setting \texttt{HardQCD:all = on}. To have better efficiency in the event selection, we set \texttt{PhaseSpace:pTHatMin= 100}, which puts a lower bound of 100 GeV on the transverse momentum on each of the two final state legs in the $2\to 2$ processes for all classes. We generate 300k events passing the selection criteria for each class, which are used to create three binary classification datasets for each of the three signals vs the background. The leading jet's constituents in these events are used to construct jet graphs. We use $R_0=0.5$ to construct the radius graph and extract $\mathbf{h}^{(0)}_i=\Delta R_{iJ}$, and $\mathbf{x}^{(0)}_i=(\Delta y_{iJ},\Delta \phi_{iJ})$  along with $z_i$ and $\omega^{(\mathcal{N}[i])}_j$. For the ordinary EMPN architecture, we use $\mathbf{H}^{(0)}_i=\mathbf{x}^{(0)}_i$.  Since all three networks process the node features to extract relevant edge features as inputs to the learnable functions, no edge features are added. The combined datasets of the signal and background (consisting of 600k samples) are segregated into 60\% training, 20\% testing, and 20\% validation datasets.  
%---------------------
\begin{figure}[!t]
	\centering
	\includegraphics[width=0.33\linewidth]{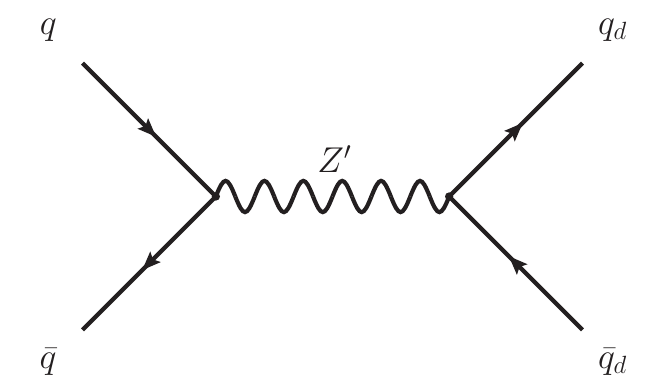}\hfill
	\includegraphics[width=0.33\linewidth]{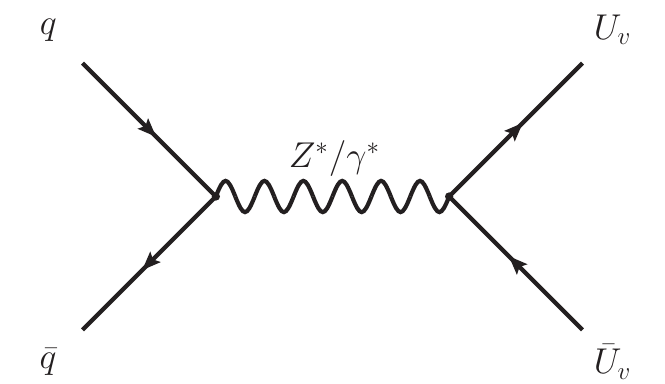}\hfill
	\includegraphics[width=0.25\linewidth]{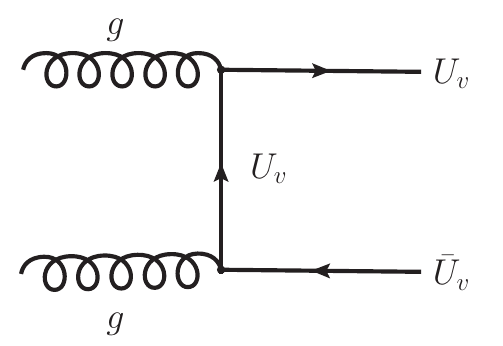}
	\caption{Representative Feynman diagrams for the dark shower event generation processes.} 
\end{figure}
%---------------------
%%%%%%%%%%%%%%%%%%%%%%%%%%%
\subsection{Network Architecture and training}
%%%%%%%%%%%%%%%%%%%%%%%%%%%
The network analysis is carried out using the \textsc{PyTorch-Geometric} \cite{fey2019fast} package. We consider an IRC-safe \texttt{EdgeConv} operation modified with the radius graph and the energy-weighted structure as an example of a non-equivariant architecture.
The message passing operation for this architecture follows from Eq. (\ref{eq:ewmp}), with $\hat{\Phi}$ being an MLP which takes the input of the form:

\begin{equation}
	\label{eq:ec_ewmp}
	\mathbf{H}^{(l+1)}_i=\sum_{j\in\mathcal{N}[i]}\;\omega^{(\mathcal{N}[i])}_{j}\;\;\hat{\Phi}^{(l+1)}(\mathbf{H}^{(l)}_i,\mathbf{H}^{(l)}_j - \mathbf{H}^{(l)}_i)\quad.
\end{equation} 

For ease of reading, we name the three architecture as \et-EMPN, \ot-EMPN, and EdgeConv-EMPN. As one of the hallmarks of equivariant architectures is the efficiency of feature extraction with a low number of tunable parameters, we consider a small variant containing about 4k parameters and a large variant containing about 150k parameters. The \et-EMPN and \ot-EMPN have the same structure of MLPs, and the difference arises from the edge input to $\Phi^{(l)}_e$ and the aggregation of the vectors. Since the EdgeConv-EMPN has only one MLP, while the equivariant ones have three, we reduce the dimensionality of the hidden (scalar) node representations in the latter to keep the number of parameters comparable.  We apply the message-passing operation twice for all architectures and extract the IRC-safe graph representation after each operation to feed to a classifier MLP. A schematic diagram for the equivariant ones is shown on the right side of figure~\ref{fig:o2_architecture}.

For the small parameter size, the equivariant architectures have a 12-dimensional updated scalar node feature after each message operation, while the EdgeConv-EMPN has 16-dimensional updated node features. For $l\in\{1,2\}$, the MLPs $\Phi^{(l)}_e$, $\Phi^{(l)}_x$, and $\Phi_h^{(l)}$ have two hidden layers with the same dimension as the updated scalar feature dimension. Similarly, the edge MLPs have two hidden layers with the same dimensions as the updated node features. An analogous architecture is repeated for the large parameter case, with the equivariant ones updating 80-dimensional scalar node features and the EdgeConv-EMPN updating 128-dimensional ones. All hidden layers have \texttt{ReLU} activation. Except for $\Phi_x^{(l)}$'s, which have a one-dimensional output with \texttt{Sigmoid} activation, we do not put any output activation for the other MLPs. The graph representation for the \et-EMPN and \ot-EMPN cases is obtained from the two scalar node representations and two vector node representations.  

In total, we have three binary classification datasets and six networks, thereby giving us eighteen training cases. We train the network from random initialisation five times for each of these cases with a batch size of 300 for 200 epochs. The networks are optimised with the \textsc{Adam}~\cite{DBLP:journals/corr/KingmaB14} optimiser with an initial learning rate of $10^{-3}$. A decay-on-plateau criterion is utilised for the learning rate, triggering a decay by a factor of 0.5 if the validation loss remains stagnant for three consecutive epochs.

%%%%%%%%%%%%%%%%%%%%%%%%%%%
\subsection{Results}
%%%%%%%%%%%%%%%%%%%%%%%%%%%
%---------------------
\begin{figure*}[!t]
   \centering
   \includegraphics[width=0.45\textwidth]{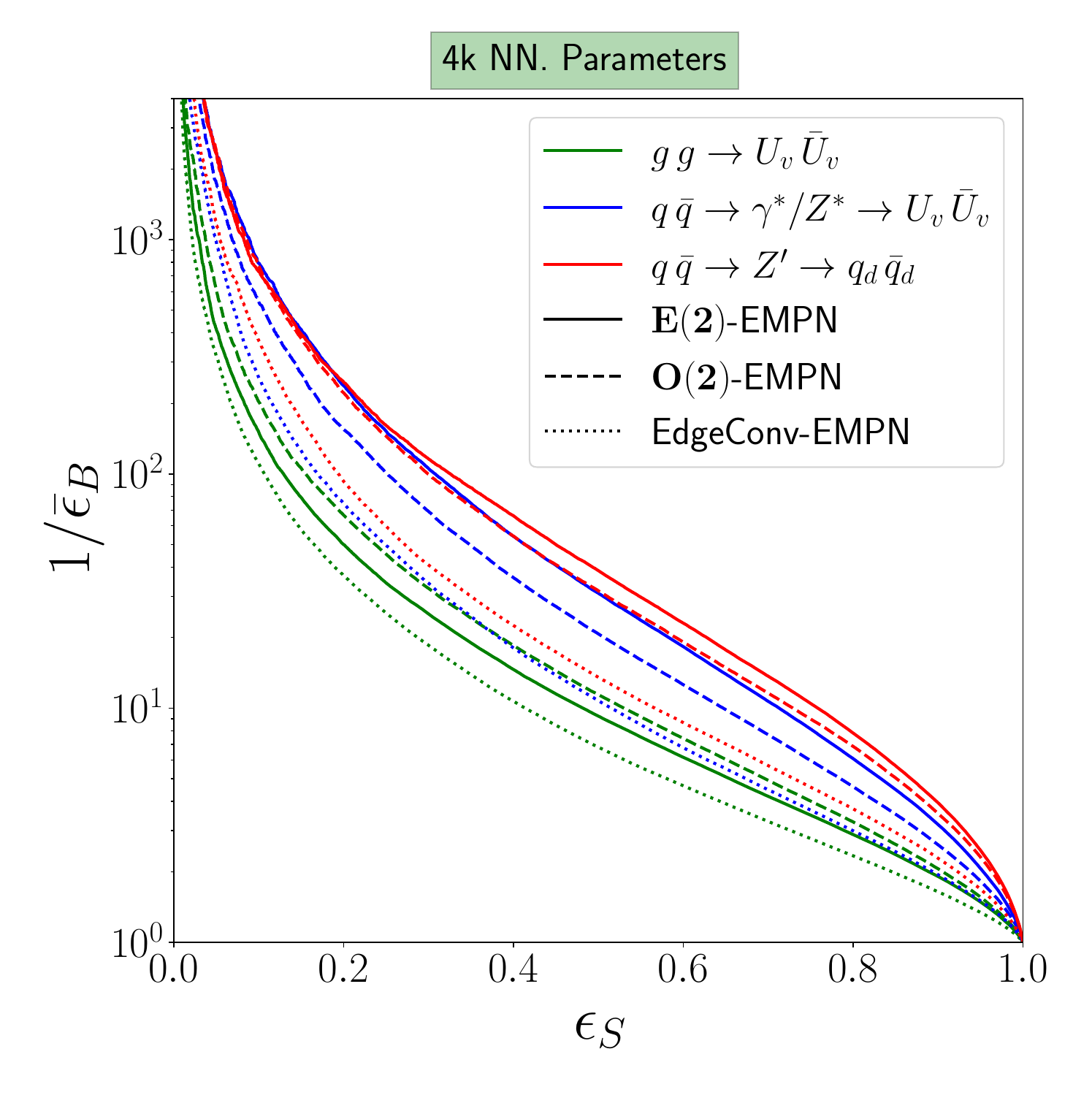}
   \includegraphics[width=0.45\textwidth]{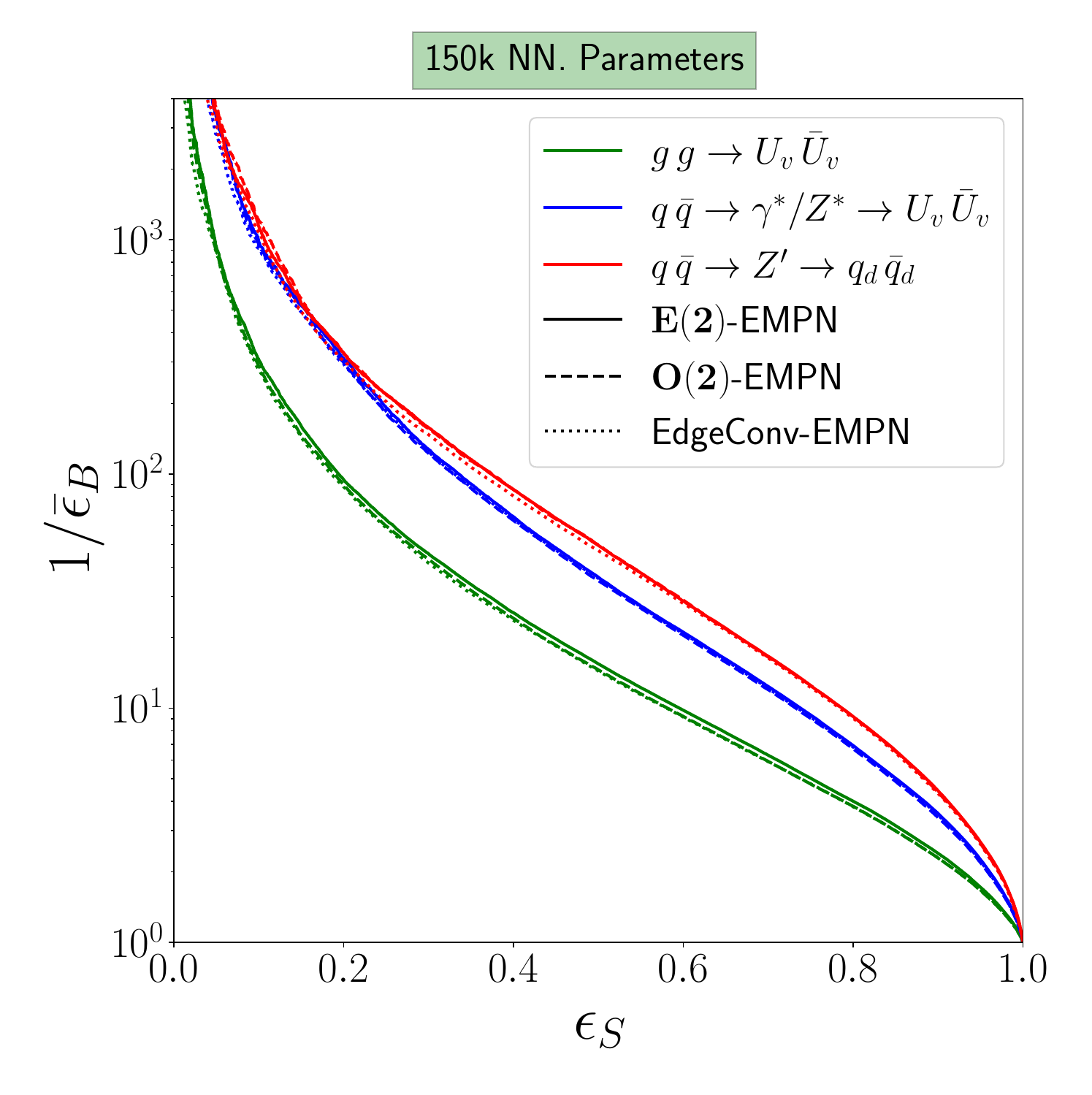}
        \caption{The network performance is shown for two scenarios: one with a low number of neural network parameters $(\approx 4k)$ and another with a high number of parameters $(\approx 150k)$, for \et-EMPN (solid), \ot-EMPN (dashed) and  EdgeConv-EMPN (dotted) for the three cases of semi-visible jet tagger.}
    \label{fig:roc}
\end{figure*}
%---------------------
We extract the prediction on the test dataset from the epoch with the lowest validation lost for each of the five training of the eighteen instances. To obtain a summary representation of the receiver-operator characteristics (ROC) curve, we extract the background acceptance $\epsilon_B$, at fixed signal efficiency $\epsilon_S$ over the five training instances of training using a custom implementation.\footnote{This was verified against the \textsc{scikit-learn}~\cite{scikit-learn} implementation %with overlapping plots 
with AUCs being equal up to four decimal points.} Extracting the mean value of the background acceptance $\bar{\epsilon}_B$, over these five instances, we plot the ROC curve between $1/\bar{\epsilon}_B$ and $\epsilon_S$ in figure~\ref{fig:roc}. The average of the area under the curve (AUC) and the standard deviation are listed in table~\ref{tab:auc}. Before comparing the relative differences for the considered architectures, we see that the classification efficiency follows the usual physics intuition. The resonant production of a heavy colour singlet $Z'$ sets a hard scale, which induces significant changes in the jets, while the non-resonant productions do not have such a hard scale. In the colour-singlet s-channel non-resonant production via the $\gamma$ and $Z$, there is no colour flow from the initial quarks to the final state (even though the $U_v$ quark carries an SM QCD charge). These jets are less QCD-like than the non-resonant production via gluons, where colour flows between the initial and final states and hence closely resembles QCD jet production.   

%---------------------
\begin{table}[!t]
%\scriptsize
    \centering
    \begin{adjustbox}{width=\textwidth}
    \begin{tabular}{|c|c|c|c|c|c|c|}
        \hline
        NN parameters & \multicolumn{3}{c}{4$k$ } & \multicolumn{3}{|c|}{150$k$} \\
        \hline
        Process & $g\, g \rightarrow U_v\,\bar{U}_v$  & $q\,\bar{q}\rightarrow \gamma^*/Z^*\rightarrow  U_v\,\bar{U}_v$ & $q\,\bar{q}\rightarrow Z'\rightarrow q_d\,\bar{q}_d$  & $g\, g \rightarrow U_v\,\bar{U}_v$  & $q\,\bar{q}\rightarrow \gamma^*/Z^*\rightarrow  U_v\,\bar{U}_v$ & $q\,\bar{q}\rightarrow Z'\rightarrow q_d\,\bar{q}_d$  \\
          \hline
        
        % \cline{2-6}
        {\et-EMPN} & 0.807±0.051	& 0.896±0.003	 &0.913±0.002	 &  0.855±0.002 & 0.906±0.001&0.923±0.001	\\
        \hline
      
       % \cline{2-6}
          {\ot-EMPN}&0.827±0.036& 0.872±0.037 &0.904±0.028 & 0.848±0.003 & 0.903±0.002&0.922±0.001 \\
        \hline
          {EdgeConv-EMPN} & 0.767±0.041& 0.817±0.042& 0.845±0.006 & 0.848±0.000 &0.904±0.000&  0.921±0.001\\
        \hline
    \end{tabular}
    \end{adjustbox}
    \caption{The network performance is illustrated for two scenarios: one with a low number of neural network parameters $(\approx 4k$) and another with a high number of parameters $(\approx 150k$), for the \et-EMPN, \ot-EMPN, and EdgeConv-EMPN. }
    \label{tab:auc}
\end{table}
%---------------------

For the large network on the right side of figure~\ref{fig:roc}, the three architectures have virtually overlapping ROCs with a very minute decrease for the non-equivariant EdgeConv-EMPN. The values of the AUCs in table~\ref{tab:auc} follow the same trend. This suggests that with sufficiently large parametrisation, the EdgeConv-EMPN can catch the distinguishing features as well as the $\mathbf{E}(2)$ and $\mathbf{O}(2)$ equivariant ones, even though the extracted features are not equivariant. In stark contrast, there is a very noticeable dip in performance for the EdgeConv-EMPN of 4k parameters for all three signal cases compared to the two equivariant architectures, with the \et-EMPN having the best classification power for the resonant and electroweak non-resonant processes, while the \ot-EMPN bettering the \et-EMPN for the non-resonant production via gluons. Moreover, the best equivariant AUCs in the low parameter case are not that far off from their corresponding best ones in the large parameter cases, with the highest difference for the hardest case of distinguishing gluon-induced production of semi-visible jets. This clearly shows that the inductive bias of equivariance in the rapidity-azimuth plane helps optimise feature extraction and reduces the need for many parameters.

%%%%%%%%%%%%%%%%%%%%%%%%%%%
\section{Summary and Conclusions} 
\label{sec:Summary}
%%%%%%%%%%%%%%%%%%%%%%%%%%%
The future roadmap of particle physics crucially depends on maximising the discovery potential of the LHC and its high luminosity phase. Recent developments in optimising searches for new physics using machine learning have demonstrated the enormous opportunity created by less traditional approaches to data analysis. Interfacing such highly adapted strategies with theoretical predictions to make robust statements about parameter measurements, whether related to SM or BSM effects, highlights infrared and collinear safety as a central theme of neural network architecture design, e.g. through energy-weighted message passing. Adding such features to the neural network design can practically increase computational demands, which requires optimisation improvements. In this work, we have considered \et~equivariance as an avenue to remove computational overhead. In particular, we have considered jet-substructure-based analysis to show that an \et~equivariant Graph Neural Network optimises the network's learning capability by directly exploiting the physical symmetry that the data set possesses. As most jet substructure observables deal with rotationally invariant quantities, we also considered \ot-equivariance which is a subgroup of \et. 

Such advances directly impact physics applications. To highlight this, we have considered a representative (but not exclusive) scenario of semi-visible jets. In theories underpinning this BSM signature, extracting, e.g., dark shower parameters critically depends on an IRC-safe implementation of the tagging algorithm as the model navigates between visibly hard and invisibly soft final states. This asks for the correct modelling of the QCD null hypothesis to draw theoretically consistent conclusions. The architecture presented in this work provides excellent signal vs. background discrimination with these desired properties. This opens up the possibility of employing the architectures proposed in this work as a new tool for BSM discovery in challenging QCD-dominated final states.

\subsection*{Acknowledgements}
A.B. is supported by the U.S. Department
of Energy under grant number DE-SC 0016013. C.E. is supported by the UK Science and Technology Facilities Council (STFC) under grant ST/X000605/1 and the Leverhulme Trust under RPG-2021-031. W.N. is funded by a University of Glasgow College of Science and Engineering Scholarship. V. S. N. and M.S. are supported by the STFC under grant ST/P001246/1. Parts of the computation utilised the Param Vikram-1000 High Performance Computing Cluster at the Physical Research Laboratory.  
%%%%%%%%%%%%%%%%%%%%%%%%%%%
\bibliographystyle{JHEP}
\bibliography{references}
%%%%%%%%%%%%%%%%%%%%%%%%%%%

\end{document}